 \documentclass[final,3p,times,twocolumn]{elsarticle}

 \usepackage{graphicx}

\usepackage{amssymb}

\def\Vec#1{\mbox{\boldmath $#1$}}

\journal{Physica C}

\begin{document}

\begin{frontmatter}




\title{
Surface-angle dependence of the tunneling spectroscopy in iron-based superconductors: sign-reversing $s$-wave scenarios
}

\author[Tokyo,TRIP]{Yuki Nagai}
\author[Nano,CREST]{Nobuhiko Hayashi}
\author[TRIP,CREST,CCSE]{Masahiko Machida}

\address[Tokyo]{
Department of Physics, University of Tokyo, Tokyo 113-0033, Japan
}
\address[TRIP]{JST, TRIP, Chiyoda, Tokyo, 102-0075, Japan}

\address[Nano]{
Nanoscience and Nanotechnology Research Center (N2RC),
Osaka Prefecture University, 1-2 Gakuen-cho, Sakai 599-8570, Japan
}
\address[CREST]{
CREST(JST), 4-1-8 Honcho, Kawaguchi, Saitama 332-0012, Japan
}
\address[CCSE]{
CCSE, Japan, Atomic Energy Agency, 6-9-3 Higashi-Ueno, Tokyo 110-0015, Japan
}

\begin{abstract}
We discuss the surface Andreev bound states in Fe-based superconductors
with the use of an effective five-band model
and investigate the surface-angle dependence of the tunneling spectroscopy by a quasiclassical approach
for an isotropic and an anisotropic $\pm s$-wave gap superconductivity. 
We show that information on the normal state is important for the Andreev bound state and 
its peak positions do not depend on the gap amplitude anisotropy. 
\end{abstract}

\begin{keyword}


Iron-pnictides, Andreev bound states, tunneling spectroscopy, theory
\PACS 
74.20.Rp, 
74.25.Op, 
74.25.Bt  
\end{keyword}

\end{frontmatter}


The discovery of novel Fe-based superconductors has attracted considerable attention because of 
high superconducting transition temperature \cite{Kamihara}.
A $\pm s$-wave pairing symmetry has been theoretically proposed as one of the candidates for the pairing symmetry 
in Fe-based superconductors.
The $\pm s$-wave symmetry means that the symmetry of pair potentials on each Fermi surface is $s$-wave and the 
relative phase between them is $\pi$ \cite{mazin,kuroki,NagaiNJP}.
The Fe-based superconductors are known to be multi-band systems and have multiple Fermi surfaces.

It is important for the identification of the $\pm s$-wave symmetry to 
detect the sign change in the pair potentials between Fermi surfaces. 
As demonstrated in studies of high-$T_{\rm c}$ cuprates, Andreev bound states are formed at a surface or a junction when the quasiparticles feel different signs of the pair potential before and after scattering~\cite{TanakaPRL}. 
Motivated by the expectation that one can extract the information on the relative phase through such Andreev bound states,
several theoretical studies on junctions and surfaces have been reported recently~\cite{onari}. 
Andreev bound states at zero energy have been experimentally observed as a zero-bias conductance peak (ZBCP) 
in tunneling spectroscopy for Fe-based superconductors~\cite{Yates}.

In this paper, to investigate Andreev bound states
we calculate the local density of states (LDOS) at a specular surface with the use of 
the extended Matsumoto-Shiba method for $n$-band superconductors \cite{Matsumoto,NagaiPRB}.
We  discuss the surface-angle dependence of the LDOS with the effective five-band model by Kuroki et al.\ \cite{kuroki}
and $\pm s$-wave pairing symmetry.

We consider the surface situated at $x = 0$ and the surface scattering potential $\check{U}(\Vec{r})$ written as $\check{U}(\Vec{r}) = U_{0} \delta(x) \check{\tau}_{3}$. 
Here, $\check{\tau}_{i}$ ($i=1,2,3$) denote Pauli matrices in Nambu space, $\Vec{r}$ is the position in the real space and 
we take the $x(y)$-axis perpendicular (parallel) to the surface. 
The surface is actually represented in the limit $U_{0} \rightarrow \infty$. 
We use units in which $\hbar = 1$. 
The retarded Green function $\check{G}^{R}(x,x',k_{y})$ in the present system
is obtained as
$\check{G}^{R}(x,x',k_{y})
=
\check{G}_0^{R}(x,x',k_{y})
-
\check{G}_0^{R}(x,0,k_{y})
[\check{G}_0^{R}(0,0,k_{y})]^{-1}
\check{G}_0^{R}(0,x',k_{y})$.
Assuming that intra-band pairings are dominant, 
$\check{G}_0^{R}$
can be divided into a sum of the Green functions defined on each band \cite{NagaiPRB}:
\begin{equation}
\check{G}_{0}^{R}(x,x',k_{y}) = \sum_{i} \int  \frac{d k_{x}}{2 \pi}  e^{i k_{x} (x- x')} \check{G}^{i}(k_{x},k_{y}), \label{eq:g0wa}
\end{equation}
where $i$ is the band index and
\begin{eqnarray}
\check{G}^{i} \equiv
\frac{
\left(\begin{array}{cc}(E + \lambda_{i}) \hat{M}_{i} &\Delta_{i} \hat{M}_{i} \\ 
\Delta^{\ast}_{i} \hat{M}_{i}& ( E - \lambda_{i}) \hat{M}_{i} \end{array}\right)
}{-|\Delta_{i}|^{2} + E^{2} - \lambda_{i}^{2}} , \: \: \: \: \: \: 
\label{eq:23}
\end{eqnarray}
with ${[} \hat{M}_{i} {]}_{jk} = {[} \hat{P} {]}_{ji} {[} \hat{P} {]}_{ki}^{\ast }$.
Here, $\hat{P}$ is the unitary matrix consisting of the eigenvectors that diagonalize the normal state Hamiltonian \cite{NagaiPRB}
represented with orbital basis,
and $\lambda_{i}$ ($i=1,2,\cdots,n$) denote the eigenvalues.
$\Delta_{i}$ are the superconducting pair potentials.
Then, the $k_{x}$ integration can be performed on each band independently. 
The surface LDOS at $x = 0$ is written as 
$
N(E) = - {\rm Im} \: \bigl[{\rm Tr} \: \int \frac{d k_{y}}{2 \pi} \check{G}^{R}(x=0,x'=0,k_{y}) \bigr]/\pi$.
%

%
%
First, let us consider the ZBCP in the surface LDOS.
With the use of a quasiclassical approximation procedure described in Refs.\ \cite{Matsumoto,NagaiPRB},
one can obtain the appearance condition of the ZBCP from the above formulation as 
\begin{eqnarray}
{\rm det}\: 
\left(\begin{array}{cc} 
-\hat{I}  & \hat{L} \\ 
\hat{L} & \hat{I}
 \end{array}\right)
 = 0, \label{eq:app}
\end{eqnarray}
where $\hat{L} \equiv -i \sum_{i \in Q,l} \hat{M}_{i}(k_{Fx}^{i,l}) {\rm sgn}\{\Delta_{i}(k_{Fx}^{i,l})\}/ 2 |v_{Fx}^{i,l}| $ and 
$\hat{I} \equiv \sum_{i \notin Q,l} \frac{1}{2 \pi} \int \frac{d k_{x}}{\lambda_{i}(k_{x})} \hat{M}_{i}(k_{x})$, 
as defined in Ref.~\cite{NagaiPRB}. 
Eq.~(\ref{eq:app}) shows that the appearance condition {\it does not} depend on the anisotropy of the pair potentials $\Delta_{i}$ and it depends only on the signs of them. 
This result signifies that information on the normal state (i.e.,  the matrices $\hat{M}_{i}$ and the Fermi velocity $v_{Fx}^{i,l}$) is important 
for the ZBCP to appear.

\begin{figure}
  \begin{center}
    \begin{tabular}{p{32mm}p{35mm}}
      \resizebox{40mm}{!}{\includegraphics{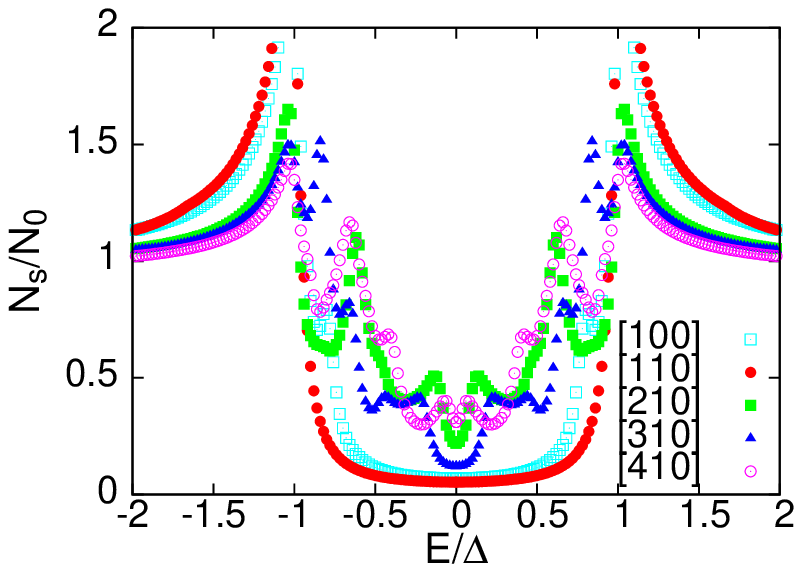}} &
      \resizebox{40mm}{!}{\includegraphics{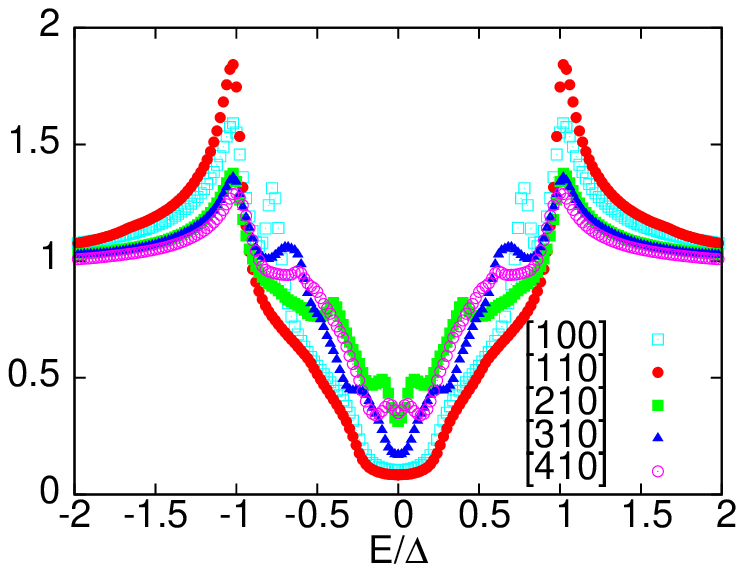}} 
    \end{tabular}
\caption{\label{fig:fig1}
Angular dependence of the surface density of states for the five-band model 
with the isotropic (Left) and the anisotropic (Right) $\pm s$-wave pair potential defined in Ref.~\cite{NagaiNJP}. 
The smearing factor is $\eta=0.05\Delta$ and the Fermi energy is $E_{\rm F} = 10.97$eV. 
($\Delta$ is the maximum gap amplitude.)
}
  \end{center}
\end{figure}

\begin{figure}
\begin{center}
\includegraphics[width = 4cm]{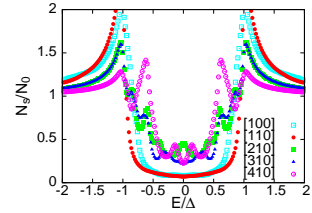}
\caption{\label{fig:fig2}
Angular dependence of the surface density of states for the five-band model 
with the isotropic $\pm s$-wave pair potential. 
$\eta=0.05\Delta$ and $E_{\rm F} = 10.94$eV. 
}
\end{center}
\end{figure}

Next, on the basis of the above general formulation~\cite{NagaiPRB},
we perform numerical calculations of the LDOS on specific Fermi surfaces obtained from
the five-band model  \cite{kuroki} of Kuroki et al.
In Figs.\ \ref{fig:fig1} ($E_{\rm F} = 10.97$eV) and \ref{fig:fig2} ($E_{\rm F} = 10.94$eV),
we show the energy dependence of the surface LDOS for various surface angle
($[nm0]$ denotes the surface normal vector). 
Comparing the left and right panels in Fig.~\ref{fig:fig1},
it appears that 
the peak positions of Andreev bound states do not depend on whether the pair potential amplitude is anisotropic or not. 
Comparison of the results for the [210] surface in 
Figs.\ \ref{fig:fig1} (left panel) and \ref{fig:fig2} indicates that the appearance condition of the ZBCP indeed depends on information on the normal state, namely it depends on the Fermi energy $E_{\rm F}$ here.
Because the peaks due to Andreev bound states do not appear for  an $s$-wave pairing without sign change,
our results also suggest that the mid-gap peaks in point-contact spectroscopy experiments 
may be the evidence of the $\pm s$-wave superconductivity.

%

In conclusion, we calculated the surface LDOS for $\pm s$-wave pair potentials with the effective five-band model. 
We showed that the peak positions do not depend on the anisotropy of the pair potential ampllitudes,
but depend on the normal-state properties.

\section*{Acknowledgments}
We thank Y. Kato, N. Nakai, H. Nakamura, M. Okumura, Y. Ohta, C. Iniotakis, M. Sigrist, Y. Tanaka and  S. Onari for helpful discussions and comments.
Y.N.~acknowledges support 
   by Grand-in-Aid for JSPS Fellows (204840). 




\end{document}